\documentclass[twocolumn,showpacs,preprintnumbers,amsmath,amssymb, superscriptaddress, prb]{revtex4-1}

\usepackage{stmaryrd}
\usepackage{txfonts}
\usepackage{amssymb}
\usepackage{mathrsfs}
\usepackage{graphicx}
\usepackage{dcolumn}
\usepackage{bm}
\usepackage{epsfig}
\usepackage{color}                    
\usepackage{hyperref}                 

\begin{document}

\preprint{\href{http://link.aps.org/doi/10.1103/PhysRevB.87.100508}{S. Z. Lin and C. Reichhardt, Phys. Rev.B {\bf 87}, 100508(R) (2013).}}

\title{Stabilizing fractional vortices in multiband superconductors with periodic pinning arrays}
\author{Shi-Zeng Lin}
\affiliation{Theoretical Division, Los Alamos National Laboratory, Los Alamos, New Mexico 87545, USA}

\author{Charles Reichhardt}
\affiliation{Theoretical Division, Los Alamos National Laboratory, Los Alamos, New Mexico 87545, USA}

\begin{abstract}
Multiband superconductors support the excitation of vortices with fractional quantum flux, known as fractional vortices. In the ground state, the fractional vortices in different bands bond together to form a composite vortex with the standard flux quantum $\Phi_0$; thus, it is difficult to stabilize the fractional vortices. Here we show that fractional vortex lattices can be stabilized in multiband superconductors with a periodic pinning array at the half matching field of the 
composite system, when full matching of the fractionalized vortices occurs. 
In the presence of a high current, the fractional vortices in different bands decouple and move at different velocities. 
When the current is turned off suddenly, the fractional vortices in different bands may be trapped at different pinning sites. 
This system also exhibits rich dynamic behavior, and 
for fractional vortices in two-band superconductors we find
a phases where the vortices  one band are pinned and the vortices 
in the other band are moving. 
These different phases can be 
observed in transport measurements, and produce a new type
of self-induced Shapiro steps that can arise 
under the application of only a dc drive.
\end{abstract}


\date{\today}

\maketitle
Multiband superconductivity continues to attract 
enormous attention as more and more multiband superconductors are discovered, such as $\rm{MgB_2}$ \cite{Nagamatsu01} and iron-based superconductors \cite{Kamihara08}. Multiband superconductors have unique features that cannot be found in single band superconductors \cite{Leggett66,Tanaka02,Babaev02,Gurevich03, Gurevich03b, Babaev05,Smorgrav05, Gurevich06, Silaev11, Komendova2012,Lin2012PRL}. Examples include fractional vortices \cite{Babaev02,Bluhm06}, short-range repulsion and long-range attraction between vortices \cite{Babaev05,Moshchalkov09}, oscillations of superconducting phase difference between different bands (the Leggett mode)\cite{Leggett66,Blumberg07} and the time-reversal symmetry broken phase \cite{Stanev10}, just to name a few. 

In multiband superconductors, electrons in each band form a superfluid condensate characterized by a complex gap function $\Delta_\mu \exp(i\theta_\mu)$. It is possible that the phase $\theta_\mu$ of the different condensates can wind by $2l_j\pi$ with an integer $l_j$. If the phase of the $j$-th band winds by $2l_j \pi$, it carries fractional flux $\Phi_j=l_j\frac{\Delta_j^2}{m_j}\left[\sum_\mu{\Delta_\mu^2}/{m_\mu}\right]^{-1}\Phi_0$, where $m_j$ is the electron's effective mass and $\Phi_0=hc/2e$ is the flux quantum \cite{Babaev02}.  When the winding numbers 
in different condensates are the same, $l_j=l$, a composite vortex 
appears carrying $l\Phi_0$ flux. The composite vortex can be viewed as a bound state of fractional vortices in different bands with the vortex cores located at the same position. For multiband superconductors under a strong magnetic field, the ground state is a lattice of composite vortices with $l_j=1$. Recently there has been growing interest in the stabilization of fractional vortices in multiband superconductors.

It was shown that the energy of the fractional vortices diverges logarithmically or linearly in bulk superconductors \cite{Babaev02}, which renders the fractional vortices unstable in bulk. This logarithmic divergence is cut off in a finite sample or in a mesoscopic superconductor. Thus the fractional vortices can be stabilized in mesoscopic superconductors, as demonstrated in Refs. \onlinecite{Chibotaru10,Pina12,Geurts10} by numerical solution of the Ginzburg-Landau equation. The fractional vortices can also be stabilized near the surface of a multiband superconductor \cite{Silaev11b}. It was demonstrated that under a driving current, the composite vortex lattice dissociates 
if the disparity of the fractional flux and coherence length in the different bands is large \cite{szlin13PRL}, 
resulting in flux flow of the decoupled fractional vortex lattices. 
The composite vortex lattice can also be decoupled by thermal fluctuations. As temperature increases, the fractional vortex lattice in the band with a weaker phase rigidity first melts while the fractional vortex lattice order in the band with a stronger phase rigidity remains \cite{Smorgrav05}.  

In this work, we show that fractional vortices can be stabilized in bulk multiband superconductors with a periodic array of pinning sites. For square periodic pinning arrays, peaks in the critical current occur when the number of vortices is an integer multiple of the number of pinning sites \cite{Baert1995,Harada1996,Martin1997,Welp2002,Karapetrov2005,Swiecicki2012,Reichhardt1998,Berdiyorov2006} as well as at fractional fields such as $1/4$, $1/2$ and $3/4$, where crystalline vortex states form that can match with the symmetry of the pinning array 
\cite{Field2002,Reichhardt2001}. We consider a square pinning array with matching field $B_{\phi}$ under an applied field creating a composite vortex lattice with $B = B_{\phi}/2$, so that when the vortices are fractionalized, the total number of vortices will equal the first matching field. 
 
To stabilize a fractional vortex lattice we apply a 
current large enough to cause the fractional vortex lattices in 
different bands to move at different velocities. 
When we suddenly turn off the current, the fractional vortices in 
different bands can be trapped by different pinning sites.  
Although the resulting fractional vortex lattice is  metastable, the lifetime is long if the pinning is strong and
the final vortex number equals the first matching field.   
In addition, we show that the   
dynamic phase diagram for vortices in multiband superconductors exhibits
a wide range of dynamical behaviors, including a phase where the vortices in one band are pinned while
the vortices in the other band are moving. We found that this system can exhibit novel self-induced Shapiro 
steps in the flux flow of fractional vortex lattices with different velocities, even though only a dc current is applied.   

We consider a two-band superconductor with interband Josephson coupling. To demonstrate our idea we adopt the phenomenological London theory, which is valid when $\lambda$ is much larger than the coherence length $\xi_\mu$ in different bands, such that the amplitude of the order parameter is approximately constant in space. The London free energy functional density is
\begin{equation}\label{eq1}
\mathcal{F}_L=\frac{1}{8\pi }\sum _{\mu}\left[\frac{1}{\lambda _{\mu}^2}\left(\mathbf{A}-\frac{\Phi _0}{2\pi }\nabla \theta _{\mu}\right)^2+(\nabla \times \mathbf{A})^2\right]-\gamma\cos \left(\theta _1-\theta _2\right)
\end{equation}
where $\lambda_{\mu}=\sqrt{(m_{\mu} c^2)/(4\pi n_{\mu} e^2)}$ is the London penetration depth for each condensate with superfluid density $n_{\mu}$. 
$\mathbf{A}$ is the vector potential and $\gamma$ is the interband Josephson coupling. The effective penetration depth for the two-band system is $\lambda^{-2}=\sum_{\mu=1}^2 \lambda_{\mu}^{-2}$. For $\rm{MgB_2}$, $\gamma>0$; thus, in the ground state, $\theta_1=\theta_2$. For certain iron-based superconductors, $\gamma<0$; thus, in the ground state $\theta_1=\theta_2+\pi$.\cite{Mazin08,Kuroki08} In the proposed liquid hydrogen superconductor,\cite{Ashcroft68,Jaffe81} the interband Josephson coupling is absent \cite{Moulopoulos1999}, $\gamma=0$.

 \begin{figure}[b]
\psfig{figure=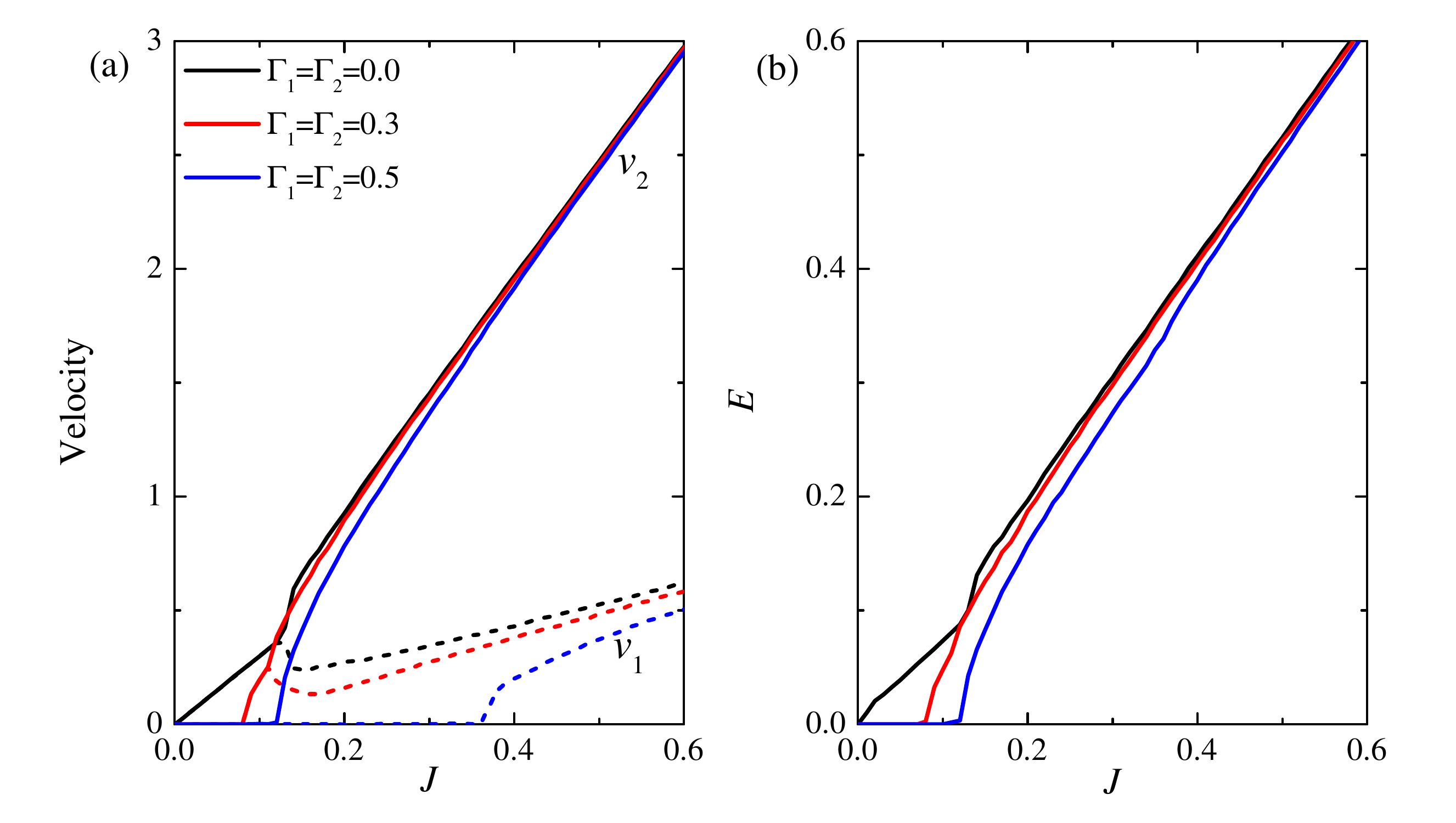,width=\columnwidth}
\caption{\label{f0}(color online) (a) Dependence of the velocity of fractional vortices on current and (b) IV characteristics for different pinning strengths $\Gamma_\mu$. For a weak pinning strength, the two vortex lattices depin simultaneously at the depinning current and then move together with the same velocity. For a large current, they split into two moving fractional vortex lattices with distinct velocities. In the case with a stronger pinning, they depin at different currents.  Here $n_p=n_v$ and $\Phi_2/\Phi_1=5.0$. }
\end{figure}
 \begin{figure}[t]
\psfig{figure=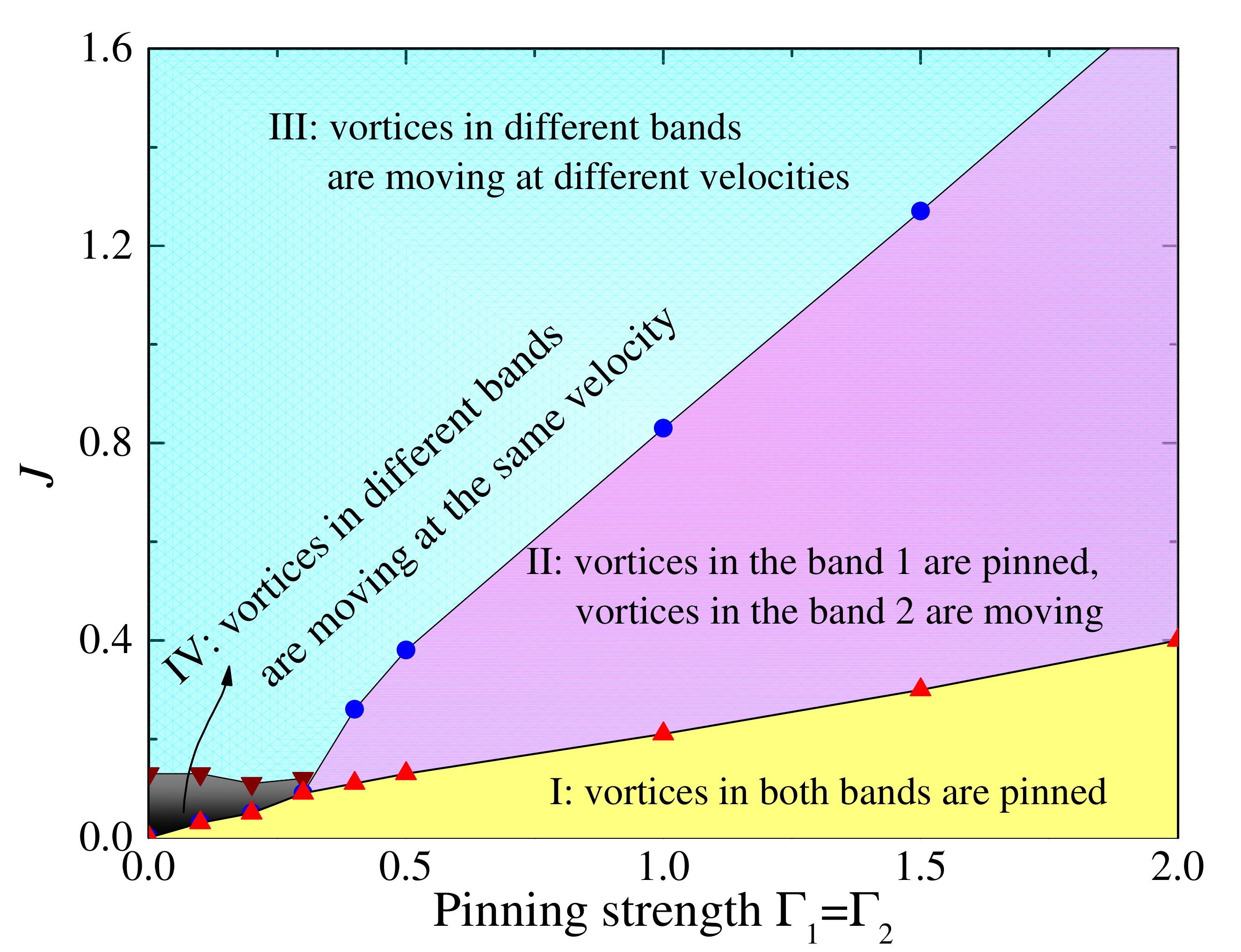,width=\columnwidth}
\caption{\label{f0b}(color online) (a) Dynamic phase diagram for two-band superconductors with a pinning array subject to a dc current. Four different regions can be identified. Here $n_p=n_v$ and $\Phi_2/\Phi_1=5.0$. }
\end{figure}

In the vortex lattice phase, the profile of the superconducting phase $\theta_\mu$ is given by the following equation obtained by minimizing Eq. (\ref{eq1}) with respect to $\theta_\mu$,
\begin{equation}\label{eq2}
\frac{\Phi _1\Phi _2}{16\pi ^3\lambda ^2}\nabla ^2\left(\theta _1-\theta _2\right)-\gamma  \sin \left(\theta _1-\theta _2\right)=0,
\end{equation}
subject to the boundary condition $\nabla\times(\nabla \theta_{\mu})=2\pi\sum_{\mu, j}\delta(\mathbf{r}-\mathbf{r}_{\mu, j})$, where $\Phi _{\mu}=\lambda ^2\Phi _0/\lambda _{\mu}^2$ is the fractional quantum flux and $\mathbf{r}_{\mu, j}$ is the core position of the fractional vortex in the band $\mu$. We assume that the vortex lines are straight. At higher magnetic fields the Josephson coupling becomes negligible \cite{szlin13PRL}. Neglecting the Josephson coupling $\gamma=0$, Eq. (\ref{eq1}) can be rewritten as \cite{Silaev11b,Smiseth2005}
\begin{equation}\label{eq3}
\mathcal{F}=\frac{1}{8\pi }\left[\mathbf{B}^2+\lambda ^2(\nabla \times \mathbf{B})^2\right]+\frac{\Phi _1\Phi _2}{32\pi ^3\lambda ^2}\left[\nabla \left(\theta _1-\theta _2\right)\right]^2.
\end{equation}
The energy only involves quadratic terms; thus, Eq. (\ref{eq3}) describes the pairwise interaction between vortices. The first term is the electromagnetic coupling between vortices and the second term accounts for the coupling between the superconducting phases. The electromagnetic coupling contributes to the repulsive interaction between fractional vortices in different bands and in the same band. The term proportional to $(\nabla\theta_\mu)^2$ describes the repulsive interaction between vortices in the same band, and the cross term proportional to $-\nabla\theta_1\nabla\theta_2$ describes the attraction between vortices in different bands. This attraction outweighs the repulsion due to the magnetic coupling and the net interaction between vortices in different condensates is attractive. The repulsion between the fractional vortices in the same band $\mu$ at a separation  $\mathbf{r}_{\mu, ij}\equiv\mathbf{r}_{\mu,i}-\mathbf{r}_{\mu,j}$ thus can be written as \cite{szlin13PRL,Smiseth2005}
\begin{equation}\label{eq4}
V_{\rm{intra}}(r_{\mu, ij})=\frac{\Phi _{\mu}^2}{8\pi ^2\lambda ^2}K_0\left(\frac{r_{\mu, ij}}{\lambda }\right)-\frac{\Phi _1\Phi _2}{8\pi ^2\lambda ^2}\ln \left(r_{\mu, ij} \right),
\end{equation}
and the attraction between two vortices in the different condensates with a distance $\mathbf{r}_{12, ij}\equiv\mathbf{r}_{1,i}-\mathbf{r}_{2,j}$ is
\begin{equation}\label{eq5}
V_{\rm{inter}}({r}_{12, ij})=\frac{\Phi _1\Phi _2}{8\pi ^2\lambda ^2}\left[K_0\left(\frac{{r}_{12, ij}}{\lambda }\right)+\ln \left({r}_{12, ij}\right)\right],
\end{equation}
where $K_0(r)$ is the modified Bessel function of the first kind. Equations (\ref{eq4}) and (\ref{eq5}) are valid away from the vortex core.

We then introduce pinning arrays into the system. The interaction between the pinning site at $\mathbf{r}_{p}$ and the fractional vortex in the $\mu$-th band at $\mathbf{r}_{\mu}$ is modeled as
\begin{equation}\label{eq6}
U_{\mu,p} \left(\mathbf{r}_{\mu}-\mathbf{r}_{p}\right)=-\Gamma_{\mu} \exp \left[-{\left(\mathbf{r}_{\mu}-\mathbf{r}_{p}\right)^2}/{l _{\mu}^2}\right],
\end{equation}
where $\Gamma_{\mu}$ characterizes the pinning strength and $l_\mu$ is the pinning range. The equation of motion for the fractional vortices in different bands thus is governed by the following overdamped dynamics
\begin{align}\label{eq7}
\eta_\mu\partial_t \mathbf{r}_{\mu, i}=-\nabla_{{r}_{\mu, i}}{ (V_{\rm{intra}}+V_{\rm{inter}}+U_{\mu,p})}+ {\mathbf{J} \Phi_{\mu}}/{c},
\end{align}
where the last term on the right-hand side of Eq. (\ref{eq7}) is the Lorentz force from a current density $\mathbf{J}$. The fractional vortices with a larger flux experience a stronger Lorentz force. $\eta_{\mu}$ is the viscosity of the fractional vortices. The viscosity depends on the size of normal core and is given by $\eta_{\mu}=\Phi_0^2/(2\pi c^2\xi_{\mu}^2)$ following the Bardeen-Stephen model \cite{TinkhamBook}. In the absence of a pinning potential $U_{\mu,p}=0$, it was shown in Ref. \onlinecite{szlin13PRL} that the two fractional vortex lattices move with the same velocity for a low current. At a high current, the two fractional vortex lattices decouple from each other resulting in a decoupled flux flow of fractional vortex lattice. In the presence of an ac current superposed on a dc current, Shapiro steps are induced. 

 \begin{figure}[t]
\psfig{figure=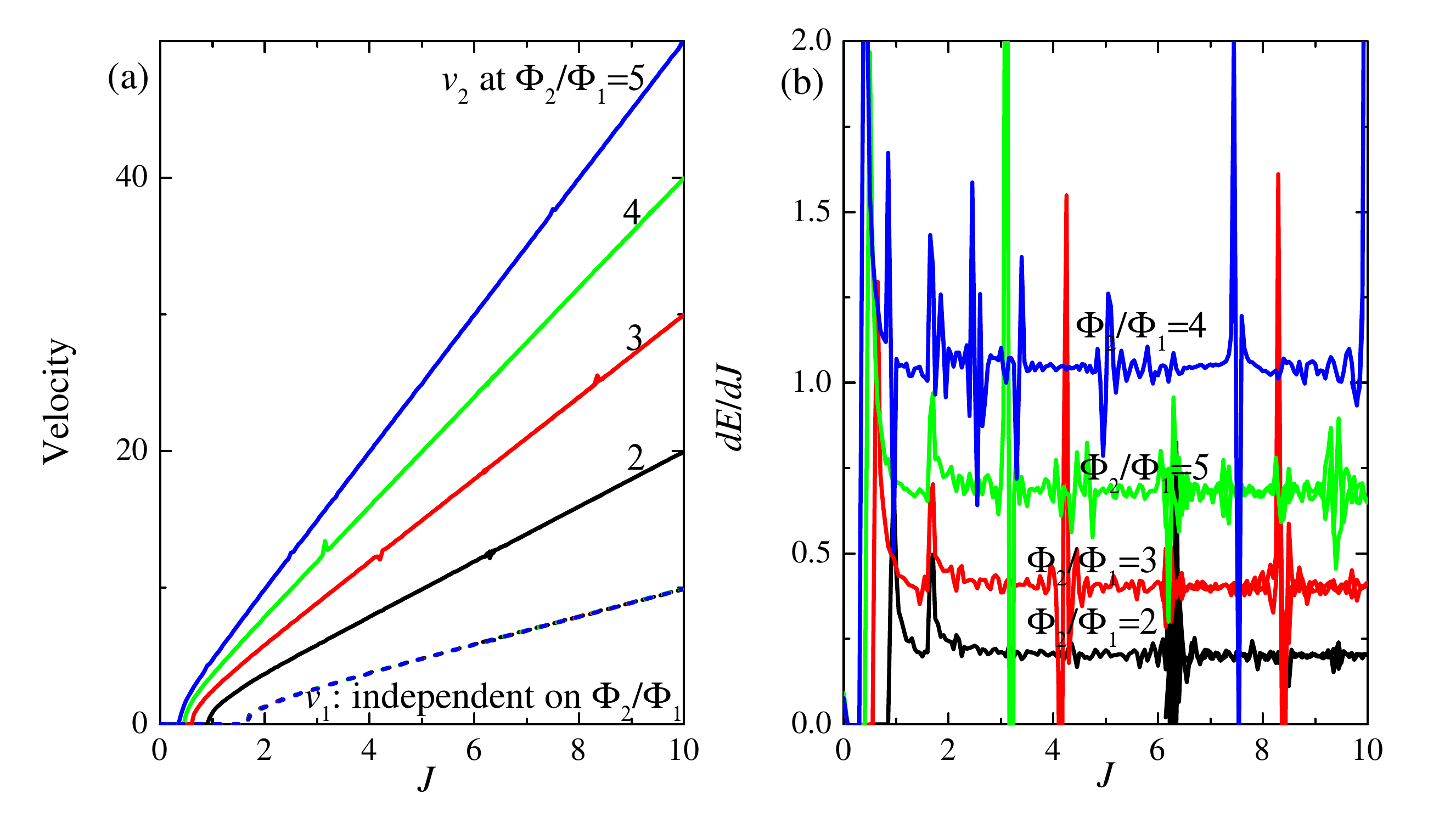,width=\columnwidth}
\caption{\label{f1}(color online) (a) Current-velocity curves and (b) the corresponding differential resistivity $dE/dJ$ for different $\Phi_2/\Phi_1$. The spikes in the velocity are due to the resonance among the two moving fractional vortex lattices and the pinning array. The first two peaks from the left in (b) are due to the depinning transition of the two composite vortex lattices. The other peaks in (b) are due to the Shapiro steps. Here $n_p=n_v$ and $\Gamma_1=\Gamma_2=2.0$.}
\end{figure}

We solve Eq. (\ref{eq7}) numerically to obtain the velocity of fractional vortices at a given $J$.  From the power balance condition $\eta_1 v_1^2+\eta_2 v_2^2= J E a^2$ with $E$ the electric field and $a$ the vortex lattice constant, we obtain the electric field. We introduce dimensionless units for force: ${\Phi _1 \Phi _2}/({8 \pi ^2 \lambda ^3})$; length: $\lambda$; time: $8 \pi ^2 \eta _1 \lambda ^4/(\Phi_1\Phi_2)$; and current: $c\Phi_2/(8 \pi ^2 \lambda ^3)$, and we set $\eta_1=\eta_2$ and $l_\mu=1$. We consider two pinning densities $n_p=n_v$ and $n_p=2n_v$, where $n_v$ is the density of the composite vortices. In the case $n_p=n_v$ we use a square pinning array with a lattice constant $a_{p,x}=a_{p,y}=5.0$. For $n_p=2n_v$, we use a rectangular pinning array with $a_{p,y}=2a_{p,x}=5.0$. The total number of composite vortices is $256$. We use a square simulation box with size $80\times 80$ and apply periodic boundary conditions. Equation (\ref{eq7}) is integrated using the second order Runge-Kutta method with a time step $\Delta t=0.02$.

The attraction between two fractional vortex lattices in different bands is a periodic function of space with a period equal to the lattice constant. The maximal attraction is $F_a={\Phi _1\Phi _2a}/({64\pi ^6\lambda ^4})$.\cite{szlin13PRL} As shown in Fig. \ref{f0} obtained numerically, in the case when the maximal pinning force $F_{\mu, p}=-\nabla U_{\mu, p}$ is much smaller than $F_a$, i.e. $F_{\mu,p}\ll F_a$, the two fractional vortex lattices depin simultaneously at a current $J_d=(F_{1,p}+F_{2, p})c/\Phi_0$. After depinning, they move with equal velocities until the current is large enough to decouple them as in the case of clean systems. When they are both pinned or move with the same velocity, they bond together to form a composite vortex lattice. Once they move at different velocities, the composite vortex lattice splits into two fractional vortex lattices. In the IV curve the voltage is zero in the pinned phase if one neglects the creep motion of vortices. A nonzero electric field is induced when the fractional vortex lattices move at the same velocity. When the composite vortex lattice splits, the differential resistivity increases, as shown in Fig. \ref{f0} (b).  In the other limit when $F_{\mu,p}\gg F_a$, the two fractional vortex lattices depin at different currents $J_{\mu,d}=F_{\mu, p}c/\Phi_\mu$, and they move at different velocities once depinned. The dynamic phase diagram is constructed in Fig. \ref{f0b}. As $\Gamma_\mu$ increases from $0$, the depinning current increases. Meanwhile the region of flux flow of the composite vortex lattice shrinks and finally disappears. Then the two fractional vortex lattices depin separately. At large current $J\gg 1$, the two fractional vortex lattices move at different velocities $v_\mu\approx J\Phi_\mu/(c\eta_\mu)$. All these dynamical phase transitions appear as features in the IV curve.

 \begin{figure}[t]
\psfig{figure=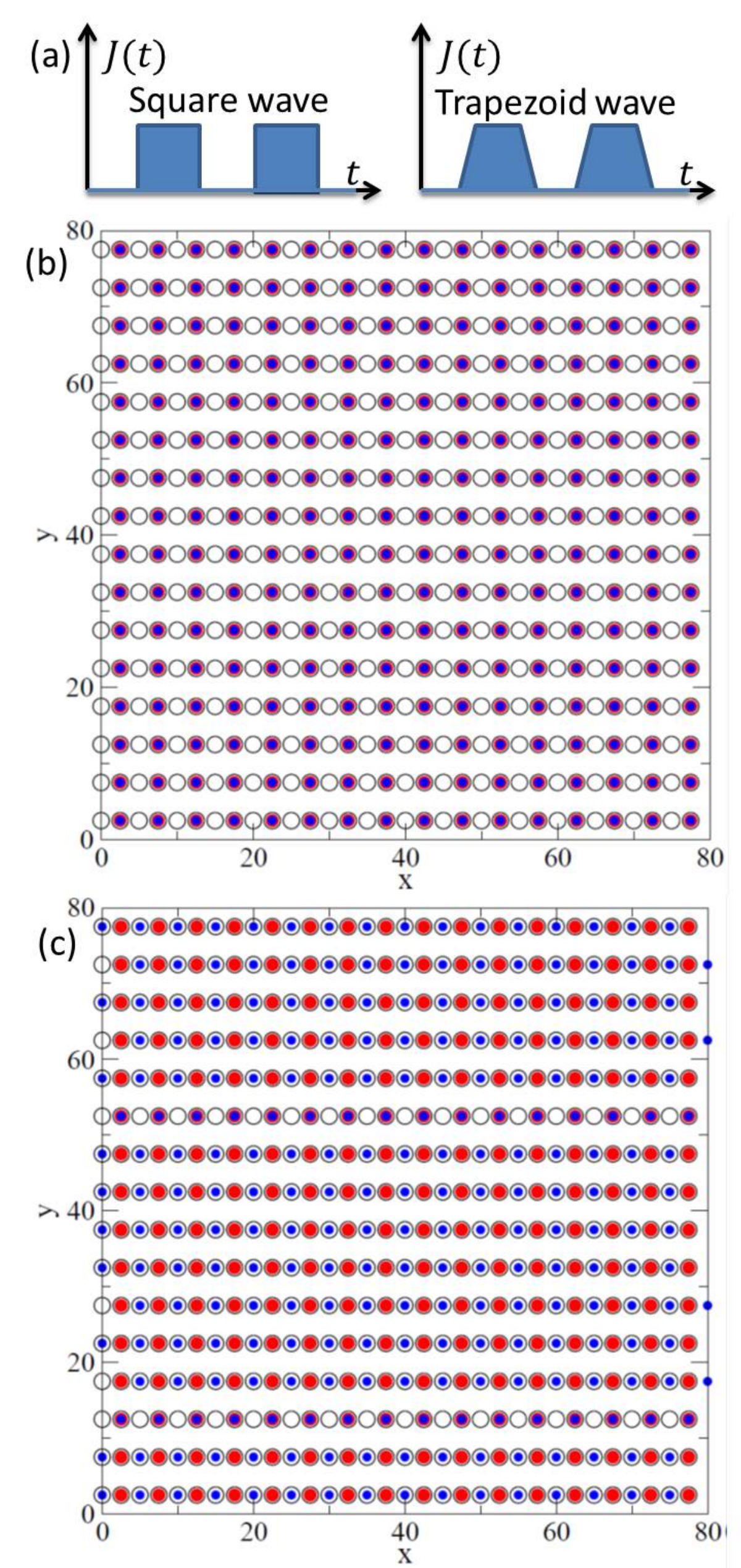,width=8.0cm}
\caption{\label{f3}(color online) (a) Schematic view of the periodic current drive used in the simulations. Snapshot of the vortex configuration (b) in the ground state and (c) after a current quench. Open circles are pinning sites; blue and red circles represent the fractional vortices in different bands. Some fractional vortices are trapped at different pinning sites after the current quench. Here $n_p=2n_v$, $\Gamma_1=\Gamma_2=2.0$, $l_1^2=l_2^2=0.5$, $J_1=3.2$,  $\Phi_2/\Phi_1=5.0$ and $\tau=100$.}
\end{figure}

When both fractional vortex lattices in bands $1$ and $2$ are moving with distinct velocities at a large $J$, the slower moving lattice in band $1$ 
experiences a periodic potential induced by the pinning array. It also feels a periodic force due to the fast moving lattice in band $2$.
 Resonance between these two periodic drivings occurs when the periods 
satisfy $l_1 T_1=l_2 T_2$. Here $T_1=a/v_1$ is the period of the ac force
 due to the pinning array acting on the vortex in band $1$,  and $T_2=a/(v_2-v_1)$ is the period of the ac force due to the fast moving lattice in band $2$.
 We have assumed that the lattice constants of the pinning array and 
the vortex lattice are equal. In terms of $v_\mu$ the resonance condition becomes $(l_2+l_1)v_1=l_2 v_2$. 
The resonance manifests as self-induced Shapiro steps in the IV curve. In the numerical simulations, we observe the resonance as shown in Fig. \ref{f1}. 
In velocity or electric field curves, 
we observe spikes as a consequence of the resonance whenever $v_\mu$ satisfies the resonance condition. 
The resonance can be seen more clearly in the differential 
resistivity $dE/dJ$ shown in Fig. \ref{f1} (b). These self-induced Shapiro steps are distinct  
from the typically observed Shapiro steps \cite{Shapiro1963} in that   
they occur without the application of an external ac drive. 
The presence of self-induced Shapiro steps 
is thus an experimentally observable 
signature of the decoupling transition between two vortex lattices. 

So far we have focused on the commensurate pinning 
case $n_p=n_v$, where the pinned  bound vortex lattice forms a square 
arrangement. 
To optimally stabilize fractional vortices, 
we consider the case $n_p=2n_v$, where in the ground state 
the composite vortices are located in a checkerboard pattern with
every other pinning site filled, as shown in Fig.~4(a).
We apply a current large enough to cause the fractional vortex 
lattices in bands $1$ and $2$ to move at different velocities. 
We then suddenly turn off the current, and find that 
most of the fractional vortex in bands $1$ and $2$ become 
trapped at different pinning sites, forming a state in which
every pinning site is occupied by a vortex. 
Thus, another square lattice forms that now has one-to-one matching 
with the pinning array.  
In this way, one may realize static fractional vortices that could be 
measured experimentally using various imaging techniques such as a SQUID. 
The resulting fractional vortices are in a metastable state; however, 
the lifetime can be long for strong pinning, and is further lengthened
due to the commensuration between the vortex configuration and the pinning
sites, which reduces the effects of fluctuations from 
vortex-vortex interactions.  In simulations, we apply a periodic square wave 
current with $J(t)=J_1$ for $0<t<\tau$ and $J(t)=0$ for $\tau<t<2\tau$. When $J=0$, 
we take a snapshot of the system, as shown in Fig.4(b), which indicates that most of the fractional vortices are trapped at different pinning sites. We have also performed simulations with other fillings such as $n_p/n_v=3$ and $n_p/n_v=4$, or using a periodic trapezoid wave form for the driving current $J$. All the results are qualitatively the same as those in Fig. \ref{f3}.

Let us discuss the possible application of the present theory to the stabilization of a half-quantum vortex in the $p$-wave superconductors, which have the potential for topological quantum computing because they host the Majorana fermion. For a small ratio of the spin superfluid density to superfluid density $\rho_{sp}/\rho_s<1$, it was shown that the half-quantum vortex can be stabilized in a mesoscopic $p$-wave superconductor \cite{Chung07}. Later this half-quantum vortex was observed experimentally \cite{Jang2011} in $\rm{Sr_2RuO_4}$, which is a strong $p$-wave superconductor candidate. It was then proposed that half-quantum vortices may be stabilized in the ground state of perforated superconducting films, which can be observed by magnetoresistance measurement \cite{Vakaryuk11}. For a large $\rho_{sp}/\rho_s>1$ where the two-half quantum vortices form a composite vortex by overlapping their normal cores, it is difficult to split them by the application of an external current as proposed in this work, because the two-half quantum vortices have the same flux and coherence length, and thus they respond to the external current in the same way. 

The creation of fractional vortices replies on the splitting of the composite vortex, which is possible only for a weak interband coupling. Weakly coupled multiband superconductors were found to be realized in $\rm{V_3Si}$\cite{Kogan09} and $\rm{FeSe}_{1-x}$\cite{Khasanov10}, and also possibly in $\rm{MgB_2}$. \cite{Xi08} The external field $H_a$ where the interband Josephson coupling becomes negligible is given by the condition  $\sqrt{\Phi_0/H_a}\ll\lambda_J=\sqrt{{\Phi _1\Phi _2}/({16\pi ^3\lambda ^2|\gamma|}})$. For $\rm{MgB_2}$, it requires $H_a> 4$ T at temperature $T=0$ K for typical parameters. \cite{szlin13PRL} The current where the composite vortex lattice dissociates is $J_d\approx 5\times 10^9\ \rm{A/m^2}$, which is much smaller than the depairing current. For $\rm{V_3Si}$ and $\rm{FeSe}_{1-x}$, the required field $H_a$ and the dissociation current $J_d$ are smaller. To stabilize fractional vortices in the pinning sites, the density of pinning sites should be comparable to that of the composite vortices. For $\rm{MgB_2}$, the period of pinning array is about $20$ nm. 

To summarize, we propose a method to stabilize fractional vortex lattices in multi-band superconductors with periodic pinning arrays. 
Under an applied drive, we identify several distinct dynamic phases that occur,  including a phase where vortices in both bands are pinned, a phase in which one species is pinned
and the other is moving, and a higher drive phase where both species are in motion but at different velocities. The different depinning transitions can be 
observed in transport measures such as $dV/dI$ curves. Another feature of the moving state is that this system exhibits self-induced Shapiro steps in the transport curves
that can arise without the application of an external ac drive. 
A fractional vortex lattice      
can be stabilized when the composite system is initially at the half-matching field of the periodic pinning array, so that only
every other pinning site captures a composite vortex. 
Once the system is in the moving
state, the drive is quenched to zero and a fractional vortex lattice appears in which every pinning site captures a fractional vortex. 

\vspace{2mm}

 \noindent {\it Acknowledgments --}
The authors are grateful to Lev N. Bulaevskii, Ulrich Welp and Cynthia Olson Reichhardt for useful discussions. This work was supported by the US Department of Energy, Office of Basic Energy Sciences, Division of Materials Sciences and Engineering.

%

\end{document}